\documentclass[lettersize,journal]{IEEEtran}
\usepackage{amsmath,amsfonts}
\usepackage{algorithmic}
\usepackage{algorithm}
\usepackage{array}
\usepackage[caption=false,font=normalsize,labelfont=sf,textfont=sf]{subfig}
\usepackage{textcomp}
\usepackage{stfloats}
\usepackage{url}

\usepackage{multirow}
\usepackage{enumitem}
\usepackage{verbatim}
\usepackage{graphicx}
\usepackage{cite}
\usepackage{xcolor,color,colortbl}
\usepackage{scalerel}
\usepackage{tikz}
\usetikzlibrary{svg.path}
\usepackage{acronym}

\definecolor{orcidlogocol}{HTML}{A6CE39}
\tikzset{
  orcidlogo/.pic={
    \fill[orcidlogocol] svg{M256,128c0,70.7-57.3,128-128,128C57.3,256,0,198.7,0,128C0,57.3,57.3,0,128,0C198.7,0,256,57.3,256,128z};
    \fill[white] svg{M86.3,186.2H70.9V79.1h15.4v48.4V186.2z}
                 svg{M108.9,79.1h41.6c39.6,0,57,28.3,57,53.6c0,27.5-21.5,53.6-56.8,53.6h-41.8V79.1z M124.3,172.4h24.5c34.9,0,42.9-26.5,42.9-39.7c0-21.5-13.7-39.7-43.7-39.7h-23.7V172.4z}
                 svg{M88.7,56.8c0,5.5-4.5,10.1-10.1,10.1c-5.6,0-10.1-4.6-10.1-10.1c0-5.6,4.5-10.1,10.1-10.1C84.2,46.7,88.7,51.3,88.7,56.8z};
  }
}

\newcommand\orcidicon[1]{\href{https://orcid.org/#1}{\mbox{\scalerel*{
\begin{tikzpicture}[yscale=-1,transform shape]
\pic{orcidlogo};
\end{tikzpicture}
}{|}}}}

\usepackage{hyperref} 
\begin{document}

\title{Centralized and Decentralized ML-Enabled Integrated Terrestrial and Non-Terrestrial Networks}

\author{Mehmet Ali~Ayg\"{u}l \orcidicon{0000-0002-1797-8238}\,,~\IEEEmembership{Student Member,~IEEE,} Halise T\"{u}rkmen \orcidicon{0000-0002-8376-0536}\,,
Mehmet \.{I}zzet Sa\u{g}lam \orcidicon{0000-0002-4670-6960}\,, Hakan Ali Çırpan \orcidicon{0000-0002-3591-6567}\,,~\IEEEmembership{Senior Member,~IEEE,}
and~H\"{u}seyin~Arslan \orcidicon{0000-0001-9474-7372}\,,~\IEEEmembership{Fellow,~IEEE}

\thanks{This work has been submitted to the IEEE for possible publication. Copyright may be transferred without notice, after which this version may no longer be accessible.}
\thanks{M. A.~Ayg\"{u}l and H. A.~Çırpan with the Department of Electronics and Communications Engineering, Istanbul Technical University, Istanbul, 34467, Turkey. M. A.~Ayg\"{u}l is also with the Department of Research \& Development, Vestel, Manisa, 45030, Turkey (e-mails: mehmetali.aygul@ieee.org and cirpanh@itu.edu.tr).
}
\thanks{M. I. Sa\u{g}lam is with the Department of Research \& Development, Turkcell, Istanbul 34880, Turkey (e-mail:izzet.saglam@turkcell.com.tr)}
\thanks{H. T\"{u}rkmen and H.~Arslan are with the Department of Electrical and Electronics Engineering, Istanbul Medipol University, Istanbul, 34810, Turkey. H.~Arslan is also with the Department of Electrical Engineering, University of South Florida, Tampa, FL, 33620, USA (e-mails: hturkmen@medipol.edu.tr and huseyinarslan@medipol.edu.tr).}}

\maketitle
\begin{abstract}
Non-terrestrial networks (NTNs) are a critical enabler of the persistent connectivity vision of sixth-generation networks, as they can service areas where terrestrial infrastructure falls short. However, the integration of these networks with the terrestrial network is laden with obstacles. The dynamic nature of NTN communication scenarios and numerous variables render conventional model-based solutions computationally costly and impracticable for resource allocation, parameter optimization, and other problems. Machine learning (ML)-based solutions, thus, can perform a pivotal role due to their inherent ability to uncover the hidden patterns in time-varying, multi-dimensional data with superior performance and less complexity. Centralized ML (CML) and decentralized ML (DML), named so based on the distribution of the data and computational load, are two classes of ML that are being studied as solutions for the various complications of terrestrial and non-terrestrial networks (TNTN) integration. Both have their benefits and drawbacks under different circumstances, and it is integral to choose the appropriate ML approach for each TNTN integration issue. To this end, this paper goes over the TNTN integration architectures as given in the 3rd generation partnership project standard releases, proposing possible scenarios. Then, the capabilities and challenges of CML and DML are explored from the vantage point of these scenarios.
\end{abstract}


\section{Introduction}

\par The conclusion of \ac{5g} standardization efforts and subsequent roll-outs have impelled academic and industry stakeholders to undertake the \ac{6g} goal: persistent connectivity, or, satisfying the need for seamless, reliable, high throughput connectivity at all times and locations \cite{giordani2020toward}. This is a challenging objective for areas with limited-to-none cellular infrastructure, scenarios where high speed vehicles are involved, and extremely dense areas. \Acp{ntn}, which span satellites, \acp{uav}, and \ac{haps}, are an attractive enabler of the \ac{6g} vision due to their large coverage areas and limited reliance on terrestrial infrastructure \cite{huaweiwp}. As such, multiple entities have been tasked with determining the exigencies for effectively \ac{tntn} \cite{NTNprojects}. The \ac{3gpp} has also released multiple technical reports and specifications on \acp{ntn} and incorporating \acp{ntn} into the \ac{5g} \ac{nr} standards, starting from \Ac{rel}-15.

\par \ac{tntn} integration is a formidable task, with the typical difficulties of heterogeneity in networks further encumbered by challenges such as \ac{ntn} device/network identification, continuous positioning and mobility tracking, cell/satellite re/selection and optimization, beam management, and so on. The difficulty in \ac{tntn} integration is three fold. Firstly, the information required for the optimization, such as satellite/\ac{ue} position/mobility, channel tracking, etc., is difficult to collect or obtain at the optimization device. Secondly, the optimization problems or models themselves are highly dimensional and complex. Finally, the mobility of the \acp{ntn} and \ac{ue} devices require frequent re-optimizations, once every 8-10 minutes at least in the case of \ac{leo} satellites \cite{8mins}. This renders conventional model-based solutions impracticable, because collecting the parameters for the optimization and performing the computation takes half this time, if not more \cite{8mins}.

\par \Ac{ai} and \ac{ml} algorithms are well equipped for solving these multi-dimensional optimization problems via their inherent ability to detect complex patterns \cite{huaweiwp}. Nonetheless, these algorithms cannot be used blindly. Their performance varies based on factors such as computational complexity, amount of training data required, and the applicability of the trained model to general scenarios. Another factor is the preferred control and processing schemes: centralized or decentralized. Individual or central devices may not have the processing capability to manage the increasingly complicated computations or the data used for network optimization may not be procurable at one location. On the other hand, coordination in centralized control and processing is much easier.

\par With respect to these factors, and others discussed in this paper, \ac{ml} approaches can fall under two main categories: \ac{cml} and \ac{dml}, so-called based on the host device(s) of the data and training process. However, choosing the appropriate approach for the \ac{tntn} scenarios is still an open issue \cite{techreport}, which this paper aims to shed light on. This paper:
\begin{itemize}
    \item Goes over the use-cases and scenarios for integrated \ac{tntn} and the properties of the associated devices, classifying them into \textit{connected} and \textit{connecting devices}. 
    \item Examines the strengths and weaknesses of \ac{cml} and \ac{dml}, with respect to the \ac{3gpp} \ac{ntn} use-cases and possible future \ac{tntn} scenarios.
    \item Suggests appropriate \ac{ml} approaches  for some scenarios and architectures with sound reasoning.
\end{itemize}

\par The organization of the remainder of this paper is as follows. Section II briefly summarises \acp{ntn} architectures and use-cases, as studied by the \ac{3gpp}. Section III defines \ac{cml} and \ac{dml} and analyzes their unique traits, pros, and cons in several scenarios of integrated \acp{tntn}. Open challenges and future study items are discussed in Section IV, and finally Section V concludes the paper. 

\begin{figure*}[!t]
\centering
\resizebox{1.99\columnwidth}{!}{
\includegraphics{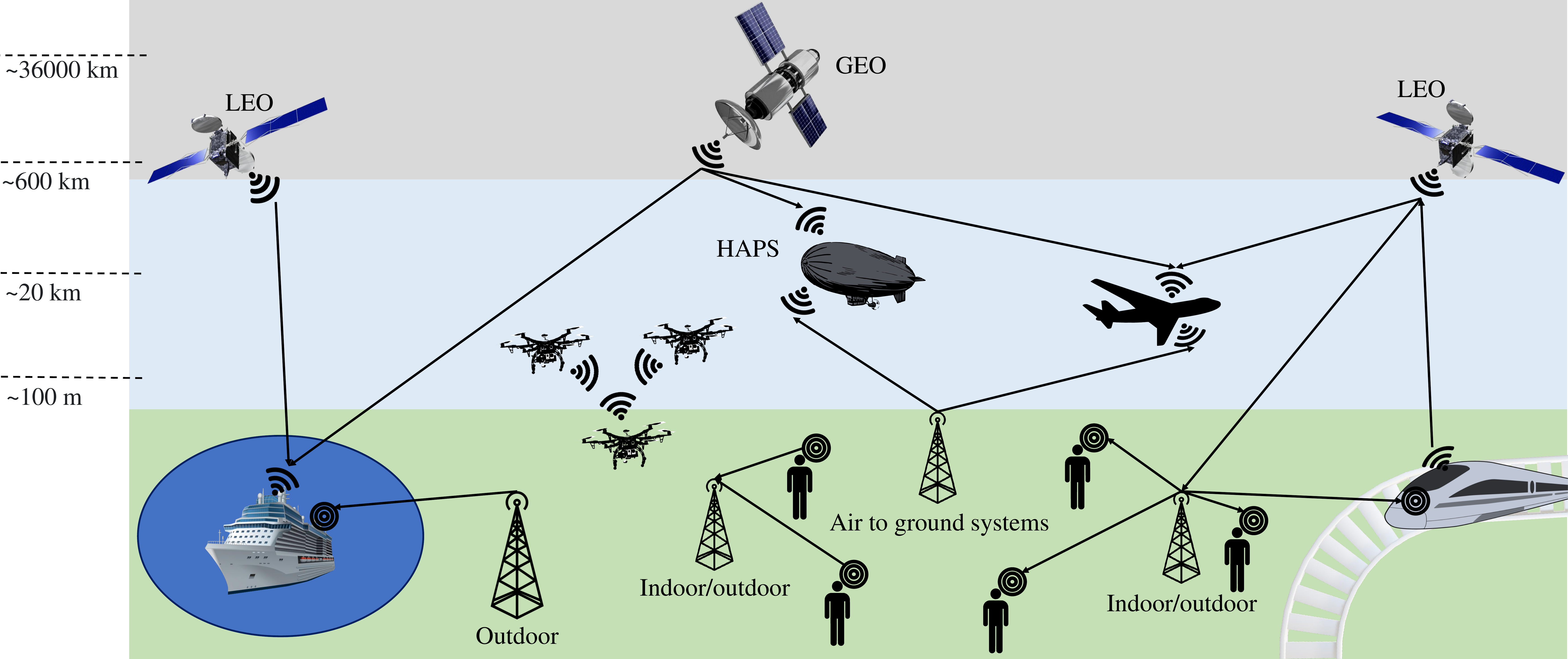}}
\caption{Use-cases of the integrated \acp{tntn}.}
\label{use_cases}
\end{figure*}

\section{NTNs in 5G NR} 
\par \Ac{ntn} systems consist of non-terrestrial devices, encompassing satellites and \ac{haps}. Their architectures include an aerial/space station that functions similarly to a terrestrial \ac{bs} or repeater, a service link between the terrestrial terminals and/or the aerial/space station, and/or a gateway that connects the non-terrestrial access network to the core network via a feeder link. The payload of the non-terrestrial device can either be transparent/bent-pipe, where frequency filtering, conversion, and amplification operations can be applied, or regenerative, where demodulation/decoding, switch/routing, and coding/modulation can be applied as well. 

\subsection{NTN Devices and UEs}

\par The devices in \acp{ntn} can be classified into six groups: satellites, \ac{haps}, low altitude aerial vehicles, maritime vehicles, terrestrial vehicles and mobile \acp{ue}, as depicted in Fig.~\ref{use_cases}. The operating and environment conditions of these devices are given in Table \ref{Table_I}. Here, the \textit{connecting devices} are essentially the non-terrestrial platforms and the \textit{connected devices} are those which are able to achieve ubiquitous connectivity through \acp{ntn}. The common issue for all users and use-case scenarios is the lack of or limited terrestrial architecture.

\begin{itemize}

    \item \textit{Satellite}: Satellites are classified as \ac{geo}, \ac{meo}, and \ac{leo}. \acp{geo} are considered stationary within a cubic volume. \acp{meo} and \acp{leo} have a fixed orbit. Several \ac{leo} satellite constellations, such as OneWeb and Starlink, intend to provide global connectivity. Delay, Doppler, strong fading effects, user mobility, and \ac{nlos} connections are concerns for connectivity.

    \item \textit{HAPS}: At a lower altitude than satellites, \ac{haps} have limited, primary \ac{tn} connections, and wide, secondary satellite connections. \ac{3gpp} has designated them as international mobile \acp{bs}. \ac{ue} positioning and trajectory planning are some concerns for connectivity. 

    \item \textit{Low altitude aerial vehicles}: These include airplanes and \acp{uav}. Limited connectivity was provided for communication with control centers in the past, but this is insufficient for the \Ac{iot} era. Frequent handovers, Doppler shifts, and dynamic channel conditions are some of the issues of connecting these devices.
    
    \item \textit{Maritime vehicles}: Maritime operations require open-sea and land-sea communication. \cite{3gpp3, ntnformaritime} considers maritime communication services as a \ac{3gpp} vertical application and a use-case of \acp{tntn} in \ac{5g} \ac{nr} networks. Large distances between \acp{ue} and unique channel conditions, such as evaporation ducts, are some concerns for connectivity.
    
    \item \textit{Terrestrial vehicles}: These are automobiles and \acp{hstv}, such as high speed trains. These vehicles are increasingly becoming more autonomous with the help of \ac{iot} devices. This requires massive number of secure and sometimes broadband connections. Doppler spread and frequent handovers is a concern for connectivity. 

    \item \textit{Mobile \ac{ue}}: The mobile \acp{ue} can be split into three sub-categories: pedestrians, \acp{hstv} passengers, and aerial vehicle passengers. Pedestrians have low-to-no mobility. \acp{ue} in automobiles can also be considered here. Connectivity is possible in dense urban and urban environments due to the presence of \ac{tn} infrastructure. However, rural or uninhabited locations necessitate alternative solutions. For aerial vehicle passengers, this, and the fact that the \ac{ue} trajectory may cross national borders further increases the complexity of the connectivity problem. Nevertheless, customers have become accustomed to continuous connectivity and expect a certain level of communication services.
\end{itemize}   

\begin{table*}[!t]
\centering
\caption{Device categories in \acp{ntn} use-cases. }
\resizebox{0.99\linewidth}{!}{
\begin{tabular}{|ccccc|}
\hline
\multicolumn{1}{|c|}{\textbf{Device Category}}        & \multicolumn{1}{c|}{\textbf{Environment Conditions}}                                                                                                                                                                                                                          & \multicolumn{1}{c|}{\textbf{Altitude}}                                                                                     & \multicolumn{1}{c|}{\textbf{Mobility Level}}                                                                                                                      & \textbf{Access to TN}                                                                          \\ \hline
\multicolumn{5}{|c|}{The ``Connecting" Devices}                                                                                                                                                                                                                                                                                                                                                                                                                                                                                                                                                                                                                                                                                                    \\ \hline
\multicolumn{1}{|c|}{Satellite}                       & \multicolumn{1}{c|}{\begin{tabular}[c]{@{}c@{}}- Free space with atmospheric and scintillation losses \\ - Satellite-satellite communication: \Ac{los} \\ - Satellite-\Ac{ue} (urban-dense urban): \Ac{nlos} \\ - Satellite-\Acp{ue} (suburban-rural): \Ac{los}\end{tabular}} & \multicolumn{1}{c|}{\begin{tabular}[c]{@{}c@{}}- \Ac{geo}: 35786 km \\ - \Ac{meo}: 7000-20000 km \\ - \Ac{leo}: 600-1200 km\end{tabular}} & \multicolumn{1}{c|}{\begin{tabular}[c]{@{}c@{}}- \Ac{geo}: Low, within a cube of 50-100 km \\ - \Ac{meo}: High, circular orbit \\ - \Ac{leo}: High, circular orbit\end{tabular}} & Yes                                                                                            \\ \hline
\multicolumn{1}{|c|}{\Acp{haps}}      & \multicolumn{1}{c|}{\begin{tabular}[c]{@{}c@{}}- Varies depending on altitude and environment \\ - Free space with AWGN, 2-tap model \\ - \Acp{haps}-\Acp{haps}/satellite communication: \Ac{los} \\ - \Acp{haps}-\ac{ue}: \Ac{los}/\Ac{nlos}\end{tabular}}                                                                    & \multicolumn{1}{c|}{8 km - 50 km}                                                                                          & \multicolumn{1}{c|}{Low}                                                                                                                                          & Yes                                                                                            \\ \hline
\multicolumn{5}{|c|}{The ``Unconnected" Devices}                                                                                                                                                                                                                                                                                                                                                                                                                                                                                                                                                                                                                                                                                                   \\ \hline
\multicolumn{1}{|c|}{Low altitude aerial vehicles}    & \multicolumn{1}{c|}{\begin{tabular}[c]{@{}c@{}}- Free space \\ - Aircraft-aircraft: \Ac{los} \\ - Aircraft-satellite/\Acp{haps}: \Ac{los} \\ - Aircraft-terrestrial device: \Ac{nlos}\end{tabular}}                                                                                                           & \multicolumn{1}{c|}{0.1 km - 6 km}                                                                                         & \multicolumn{1}{c|}{High, known trajectory}                                                                                                                       & No                                                                                             \\ \hline
\multicolumn{1}{|c|}{Maritime vehicles}               & \multicolumn{1}{c|}{- Free space with ocean clutter and evaporation duct}                                                                                                                                                                                                                                 & \multicolumn{1}{c|}{Sea level}                                                                                             & \multicolumn{1}{c|}{Low}                                                                                                                                          & Limited                                                                                             \\ \hline
\multicolumn{1}{|c|}{Terrestrial vehicle}             & \multicolumn{1}{c|}{\begin{tabular}[c]{@{}c@{}}- Varies depending on environment \\ - Urban/dense urban: \Ac{nlos} \\ - Suburban/rural: \Ac{los}\end{tabular}}                                                                                                                                     & \multicolumn{1}{c|}{Terrestrial}                                                                                           & \multicolumn{1}{c|}{\begin{tabular}[c]{@{}c@{}}- Automobiles: Low\\ - \Acp{hstv}: High\end{tabular}}                                             & \begin{tabular}[c]{@{}c@{}}- Automobiles: Low\\ - \Acp{hstv}: Limited\end{tabular} \\ \hline
\multicolumn{1}{|c|}{Mobile \ac{ue}} & \multicolumn{1}{c|}{\begin{tabular}[c]{@{}c@{}}- Indoor: \Ac{nlos} \\ - Outdoor: \Ac{nlos}/\Ac{los}\end{tabular}}                                                                                                                                                                                       & \multicolumn{1}{c|}{Terrestrial}                                                                                           & \multicolumn{1}{c|}{Low or high}                                                                                                                                  & Yes or limited                                                                                      \\ \hline
\end{tabular}}
\label{Table_I}
\end{table*}

\subsection{Use-Cases}   

\par While the devices and their operation scenarios give insight to the challenges, the use-cases effectively help to determine the requirements pertaining to communication. With the \ac{tntn} users defined, the type of service these users require are explained herein and given in Fig.~\ref{use_cases_scheme}.

\begin{itemize}[leftmargin=*]

\item \textit{Connectivity}: \Acp{tn} alone are not sufficient to provide global and ubiquitous connectivity. There are three scenarios where connectivity is compromised:
\begin{enumerate}
    \item \textit{Rural/uninhabited locations}: These locations have little to none permanent residents or visitors. As such, installing infrastructure is not feasible for operators. 
    \item \textit{Extremely dense populations/crowded events}: Rallies, concerts, sports matches, and other events push the limits of cellular networks and significantly degrade the quality of service. 
    \item \textit{High mobility \acp{ue}}: These \acp{ue} are subject to constant handovers, lowering the quality of service for the \acp{ue} and adding a burden to the networks.
\end{enumerate}

These scenarios can exist in the same instance. For example, high speed passenger trains or commercial airplanes both contain a large amount of \acp{ue} and pass through locations with no or limited cellular infrastructure. Depending on the service required, the connectivity can be multi, fixed, mobile, and mobile-hybrid. 

\par Five solution categories are proposed to enable connectivity in \cite{3gpp4}. These categories are briefly described below:
\begin{itemize}
    \item \textit{Multi/resilient}: Multi-connectivity refers to having multiple connections to increase data rate or as a back-up connection for reliable communication. Here, the non-terrestrial connection can be the back-up or main connection. Resilient connectivity aims to prevent complete network outage. Therefore, the \ac{ntn} is expected to provide broadband connectivity between the \acp{ue} and the core network in outage scenarios.
    
    \item \textit{Fixed/trunking}: In fixed connectivity the \ac{ntn} will provide the only connection. Expected to be deployed in rural or ad-hoc areas, the \acp{ntn} will provide broadband connectivity between the core network and nomadic \acp{ue}. Trunking is to provide temporary \ac{5g} connectivity in disaster or emergency situations.
    
    \item \textit{Mobile cell}: Mobile cells are the solution for the third scenario where connectivity is compromised. Here, the \acp{ntn} are expected to provide broadband connectivity between the core network and \acp{ue} on board a highly mobile platform.
    
    \item \textit{Mobile-hybrid}: This is for enabling connectivity to \acp{ue} on public transport with fixed routes. \acp{ntn} will provide a back-up or auxiliary connection for routes where the \acp{tn} have or limited capacity.
    
    \item \textit{Hot-spot-on-demand}: Here, \acp{ntn} are expected to provide temporary \ac{5g} connectivity to under-served areas.
\end{itemize}

\item \textit{Broadcasting}: This encompasses direct-to-node, direct-to-mobile, and edge network delivery broadcasts. In the former, the information is transmitted to an access point, from which it is distributed to the \acp{ue} within the network. Direct-to-mobile broadcast is used to transmit information to multiple \acp{ue} simultaneously. Such a service is required for issuing alerts to the community or responders during emergency situations and global software updates. Edge-network delivery is used to offload popular content or system updates to the edge nodes for caching and redistribution. Here, broadband connectivity is expected of the non-terrestrial devices.

\begin{figure}[!t]
\centering
\resizebox{0.99\columnwidth}{!}{
\includegraphics{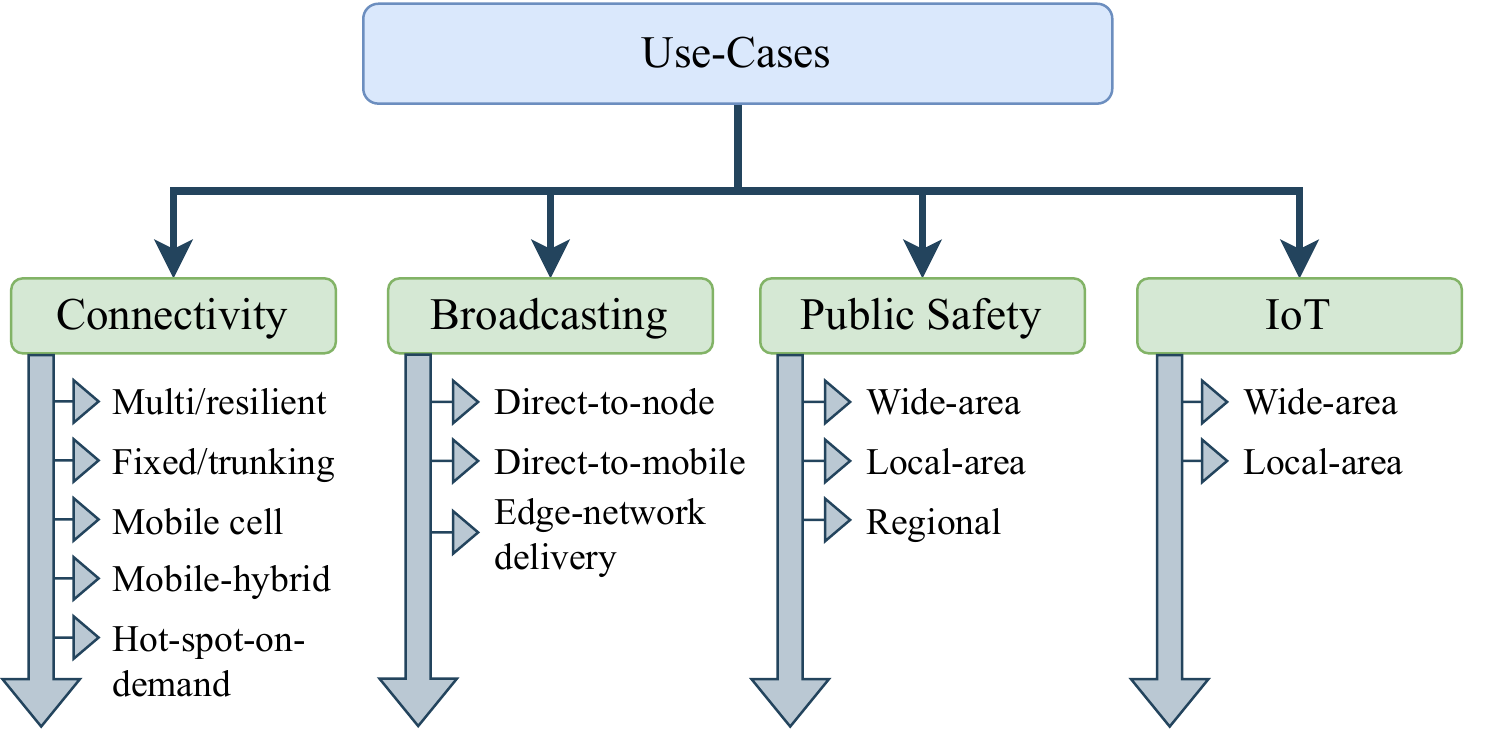}}
\caption{The illustration of the use-cases classes.}
\label{use_cases_scheme}
\end{figure}

\item \textit{Public safety}: The aim here is to provide connectivity between the emergency responders, regardless of their location and presence of terrestrial infrastructure. This can be further divided into wide-area, local, and regional public safety. The role of the \acp{ntn} would be to provide connectivity between the emergency responder \acp{ue}, tactical cells, and/or the \ac{5g} core network.

\item \textit{IoT service}: Depending on the mobility and coverage area, these use-cases can be divided into wide-area and local \ac{iot} service connectivity. \ac{iot} devices on \acp{hstv} and other scenarios involving mobility over a known area are expected to be supported by \acp{ntn} providing wide-area connectivity. Others, like devices on smart grid, are expected to be supported by \acp{ntn} providing local area connectivity. Here, the \ac{ntn} will provide connectivity between the \ac{iot} devices, their hub/central point, and/or the core network. 
\end{itemize}

\subsection{Existing Integration Issues and 3GPP Studies for Integrated TNTN}

\par Integrated \acp{tntn} should have a flexible centralized or decentralized architecture that can manage its traffic smoothly between \acp{tn} and \acp{ntn} to enable better and more intelligent utilization of the network resources. At the minimum, \acp{ue} in idle and active modes should be able to get ubiquitous coverage worldwide without receiving a congestion rejection or service degradation.

\par Achieving these goals and ubiquitous coverage has several challenges. \Acp{ntn} have a considerably larger propagation delay than \acp{tn}, increasing service interruption time during idle mode service continuity or active mode handover. Another challenge is the number of handovers in active mode due to non-\ac{geo} satellite movements. Meaning, while the coverage area of a \ac{geo} satellite is static with fixed large spot cells, \ac{leo} coverage area is changed with time and the satellite’s ephemeris, requiring frequent updates or handovers. Additionally, the cells of the \acp{ntn} have a significant signal difference between the cell center and edge. As such, the same or similar transmission parameters cannot be used for \acp{ue} at both locations. The time delay brought on by the random access procedure is another challenge, because this procedure affects \ac{ue} connection establishment and time synchronization, while integrated \ac{tntn} needs to support high-speed \ac{ue} handover and service continuity, requiring a minimum response time in the random access procedure.

\par In this regard, \acp{ntn} have been a focus of the \ac{5g} standardization efforts by \ac{3gpp}, with related works in \Ac{rel}-15, \Ac{rel}-16, and \Ac{rel}-17. \Ac{rel}-15, started in 2017, reported the results of a feasibility study targeting the channel models and deployment scenarios. Subsequently, \Ac{rel}-16 defined the minimum changes to the present standards to integrate the essential \ac{ntn} features, while \Ac{rel}-17, completed in June 2022, focused on the transparent payload architecture with earth-fixed tracking areas and \ac{fdd} systems.

\par The \ac{3gpp} \Ac{rel}-18, \Ac{rel}-19, and \Ac{rel}-20 are the upcoming releases for \ac{5g}-Advanced and focus on fine-tuning the scenarios and the usage of \acp{ntn}. Currently, the \ac{3gpp} are working on \Ac{rel}-18, the \ac{ntn} \ac{iot} enhancement \cite{3gpp6}. This release aims to cover the integration of \acp{tn} and \acp{ntn}, throughput performance, and the optimization of the \ac{gnss} sparse usage to decrease power consumption for long-term connections. Additionally, enhanced machine type communication with minimum feature updates using \acp{tntn} is also within the scope of this release. The \Ac{rel}-18 items of study also include the integrated \acp{tntn} mobility, service continuity, and coverage enhancement. These discussions cover the potential low rate codecs performance enhancements in a link budget limited context, including voice over \ac{nr}.

\section{CML and DML for integrated TNTS}

\par Pervasive system intelligence is critical for the evolution and long-term operation of integrated \acp{tntn}. Particularly, real-time decision-making substantially enhances network performance. The \ac{ml}-based remote control allows for further investigation of fundamental and unexplored characteristics of \ac{tntn} and the creation of novel communications and networking technologies, such as new protocol designs, architectures, and advanced algorithms. Network designs can be optimized to increase spectrum access flexibility, while radio channels can be modeled efficiently. Furthermore, using \ac{ml} in \ac{tntn} enables seamless autonomous communication in the presence of channel effects, such as attenuation, fading, and interference, and \ac{tntn} integration can be without the need for prior mathematical study and modeling.

\par \Ac{cml} infrastructures are designed to meet the requirements of numerous \ac{ml} models that demand locality and persistent training. The data is collected in a powerful and robust device, which also runs the \ac{ml} algorithm. It comes with the advantages of fewer resources required on training departments, networking opportunities, reduced buddy costs of training materials, and best practices across multiple sites. Still, \ac{cml}-based systems face challenges such as data delivery costs (latency), the possibility of involving poor channel and unstable connectivity conditions, coordination and scheduling durations, and generic training results, rendering them unsuitable for real-time applications. Additionally, \ac{cml} systems require sharing of sensitive operational data, which is a privacy issue. Furthermore, there are different use-cases, devices, and user types, with various problems, scenarios, and requirements. Therefore, \ac{cml} requires coordination between different use-cases and problems.

\par \Ac{dml} allows a set of local devices to locally and collaboratively participate in the training process of a global model without having to upload their local raw data to centralized servers. Thus, they restrict the amount of data transmission across the network. This adds a privacy feature and removes the delivery time, which is the time takes for data to be prepared and delivered to a central device. However, some devices may not have the capability to process complex mathematical equations of \ac{ml}, e.g., \ac{iot} and reduced capability devices. Additionally, \ac{dml} only trains with its own dataset, which may restrict the learning capability and produces internal models, i.e.: models which cannot be utilized in general scenarios.

\par
\begin{figure}[!t]
\centering
\resizebox{0.95\columnwidth}{!}{
\includegraphics{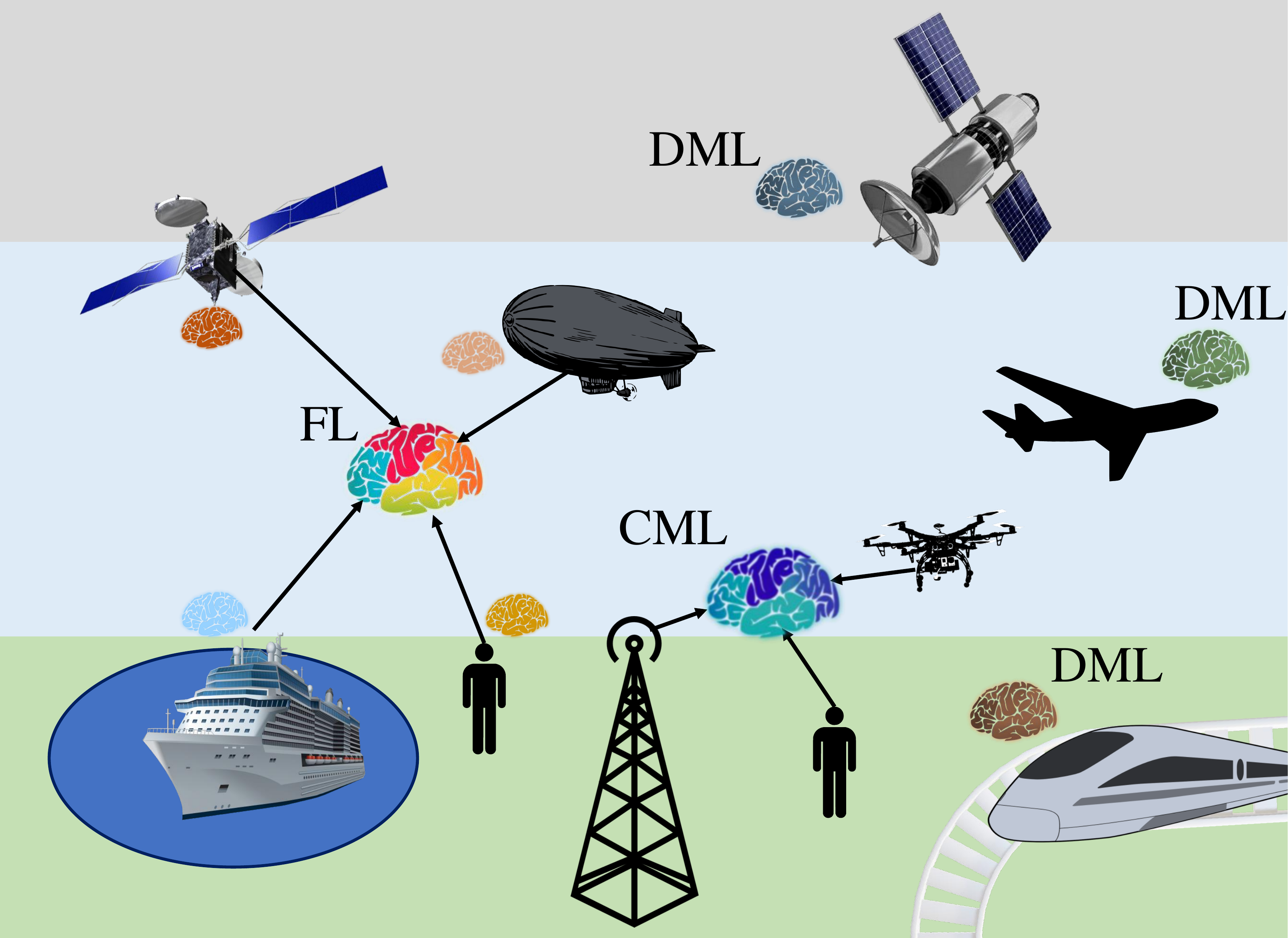}}
\caption{Example scenarios for \ac{cml}, \ac{dml}, and \ac{fl} for integrated \ac{tntn}.}
\label{cl_vs_dml}
\end{figure}

 Recently, implementation of \ac{fl}, a specialized \ac{dml} approach, has gained interest \cite{dinh2020federated}. Here, clients do local training and send their model parameters to an aggregator for further inference. This can simultaneously address the privacy issues brought by \ac{cml} techniques and the lack of generality of the models trained by \ac{dml} techniques. An illustration depicting the possible \ac{cml}, \ac{dml}, and \ac{fl} approaches are given in Fig.~\ref{cl_vs_dml}. In this figure, a \ac{leo} satellite, maritime \ac{ue}, \ac{haps} device, and a mobile \ac{ue} train their models locally (coarse learning) and share the trained parameters with the \ac{fl} cloud (brain) for fine learning. Alternatively, for the \ac{cml} approach, the \Acp{bs}, mobile \ac{ue}, and drone may send their data to a central device, where the model is trained using all the datasets and a generic model is obtained. For the \ac{dml} approach, a train (\ac{hstv}), plane, and \ac{geo} satellite can learn their individual model parameters, using the data specifically available to them.

\par In the following, we give specific scenarios and design criteria where \ac{cml} and \ac{dml} approaches are useful. However, although an approach is motivated by a specific scenario, another approach also can be useful for the scenario when it is looked at from a different perspective. Hybrid approaches, can also be useful for a scenario. 

\begin{itemize}

    \item \textit{Propagation channel and synchronization}: There are different multipath delay and Doppler spectrum models. For example, satellite communications are assumed to have outdoor and \ac{los} conditions, whereas indoor and \ac{nlos} communication conditions are assumed to be addressable using \ac{haps}. The signal is primarily direct \ac{los} for satellite-based systems and follows a Ricean distribution with a robust direct signal component; slow fading is possible due to transient signal masking, such as beneath trees and bridges. The signal in \ac{haps}-based systems comprises of considerable multipath components and follows a Ricean model. Therefore, the receiver synchronization configuration at both \ac{ue} and generation node B levels are different, including configurations such as the preamble sequence and aggregation to take into account the Doppler and specific multipath channel models, cyclic prefix to compensate the delay spread, the jitter/phase, the subcarrier spacing of the orthogonal frequency-division multiplexing signal to accommodate larger Doppler, and so on. Since the synchronization configurations are different, the same \ac{ml} algorithm may not work, therefore, a centralized learning device should be capable of doing feature extraction for all of the problems, which may be difficult in several use-cases. On the other hand, if \ac{dml} is used, each device can extract the features by itself and run its own algorithms. Therefore, the \ac{dml} approach could perform better in this scenario. 
    
    \item \textit{Cell pattern generation}: Compared to cellular networks, satellite and \ac{haps} systems often have larger, possibly mobile, cells. These cells can produce a high differential propagation delay between a \ac{ue} at the cell center and a \ac{ue} at the cell edge, particularly at low operational elevation angles. This will affect contention-based access channels when the network does not know where the \acp{ue} are. When the \acp{ue}' positions are unknown, the differential latency caused by the large cell size may cause a near-far effect during the initial access procedure. To boost performance, an extended acquisition window may be required. However, if the \ac{ue} position is known during a session, the network can correct for the differential latency. As a result, specialized signaling may be required to support these larger, mobile cells for broadcast services. Since specialized signaling is required, a \ac{dml} approach is more promising.

    \item \textit{Service continuity between \ac{tn} and \ac{ntn}}: To ensure service continuity, a handover to or from the satellite/\ac{haps} system can occur whenever a \ac{ue} leaves or enters the cellular coverage. The handover triggering mechanisms may differ depending on the circumstances, such as terminating the satellite connection as soon as there is an adequate-strength cellular signal, but only terminating the cellular connection when there is very little signal strength. The service enablers, characteristics, and measurement reports of both access technologies should be considered during the handover operation. Since there are several aspects that should be investigated jointly, \ac{cml}-based approach is promising for this issue. On the other hand, the differences in the propagation delay between \acp{ntn} and cellular networks will cause substantial jitter. If the service continuity is ensured with \ac{cml}, an extra delay time of delivery time and scheduling will be added to the system as explained before. Therefore, if the delay is important for the use-case, \ac{dml} can be preferred. Alternatively, to use the advantages of both \ac{cml} and \ac{dml}, \ac{fl} can be used for this scenario.

    \item \textit{Updating the location of the satellites}: When the locations of several \ac{leo} satellites need to be updated frequently, \ac{cml} could cause problems due to time delivery time and synchronization between different devices. Also, there can be bottlenecks and single-point failure problems for mobile \ac{leo} satellites. Therefore, a \ac{dml} approach can be promising for this problem.
    
    \item \textit{Satellite and \ac{haps}-based communication systems design}: Several design criteria exist for satellite and \ac{haps}-based communication systems. Some of these criteria are as follows.
        \begin{itemize}
            \item Maximizing throughput from the uplink \ac{ue} and the downlink satellite/\ac{haps} for a given transmit power.
            \item Maximizing service availability in cases of deep fading.
            \item Maximizing the throughput/power ratio; the operation point in the power amplifier at the satellite or the \ac{ue} should be adjusted as close to the saturation point as possible when needed. 
            \item Maximizing signal availability with slow and deep fading; vital for \ac{ue} near the cell edge, modulation and coding techniques with very low SNR operating points or other options should be studied.
            \item The \ac{mac} layer should be able to flexibly and dynamically allocate physical resource blocks to maximize spectrum efficiency and accommodate low-power terminals.
        \end{itemize}
    \noindent Since there are multiple different design criteria and they should be taken into consideration jointly, a \ac{cml} approach can yield a better performance.

    \item \textit{Terminal mobility}: Enabling communication for very high speed UEs, e.g. aircraft systems up to 1000 km/h speed \cite{3gpp2}, is a challenging task. In these speeds, \ac{cml} approach cannot work, as sharing data with other nodes may cause latency. Thus, \ac{dml} is more promising, since each node uses its own model and sharing data is not necessary.
    
    \item \textit{Security}: Integrated \acp{tntn} can include sensitive information, such as user mobility and service usage statistics or operator data. Sharing this data may not be preferred by operators or even legal, depending on the nation's data protection laws. In these cases, \ac{dml} can be the only option available.
    
    \item \textit{Dynamic service deployment}: The presence of numerous \acp{ntn}, as well as varying ground-\ac{ue} demands, could have an effect on dynamic service deployment policies. Learning their data together may increase the performance of the learning system. Therefore, \ac{cml} can be useful for these networks.
        
    \item \textit{Energy efficiency}: \Ac{cml} can be designed as a service which supports mobile network operators for energy efficiency, operational efficiency and delivering ubiquitous coverage in machine type communications. This service can help mobility and service continuity for \ac{tntn} machine type communication with easy-to-deploy, always available, secure, and reliable communications. \Ac{iot}, reduced capability devices, or sensors do not need to have a complex compute resources with this \ac{cml} service.

    \item \textit{Radio resource management adapted to network topology}: The particular cell patterns of \Acp{ntn} need to be accommodated via mobility management. Furthermore, cells in the \ac{ntn} may pass national borders. This will have an effect on cell identification, tracking and location area design, roaming and charging procedures, and location-based services in particular. Therefore, \ac{ntn} should be aware of several procedures simultaneously. This can be possible with a \ac{cml} approach. Also, the access control mechanism must respond quickly to meet fluctuating traffic demand while also taking \ac{ue} mobility requirements into account. Therefore, with the \ac{cml} both of them are learned jointly.

    \item \textit{Frequency plan and channel bandwidth}: There are several aspects for frequency plan and channel bandwidth in integrated \ac{tntn}. For example, frequency reuse and flexibility of spectrum allocation in different cells may be supported. Also, there are techniques to minimize the risk of inter-cell interference for efficient spectrum usage. To enable the targeted spectrum and the pairing between uplink/downlink bands with precise band separation, the carrier numbering can be examined. Carrier aggregation can be employed to provide equal throughput while allowing for greater flexibility in carrier allocation between cells while conforming to frequency reuse limits. These will require the system to learn relationship between different parameters and adapt the upper layers, such as \ac{mac} and network layer signaling in a specific manner. Since \ac{cml} is promising for jointly learning relationships between different problems, it may also be convenient for frequency planing.

    \end{itemize}

\section{Challenges and Study Items}

\par There are considerable challenges that need to be overcome in order to implement \ac{cml} and \ac{dml} approaches efficiently. To make integrated \ac{tntn} a reality with these \ac{ml} approaches following aspects should be studied.

\begin{itemize}

    \item \textit{Simulation analysis}: Simulation analysis should be made for \ac{cml} and \ac{dml}. These are as follows.
    
    \begin{itemize}

    \item Routing techniques with \acp{uav}, \ac{leo}, \ac{meo}, \ac{geo} satellites, and \ac{haps}, particularly in hierarchical architecture.
    
    \item Proper identification and localization and optimal trajectory design.
    
    \item Analyses to ensure privacy, integrity, and secrecy.
    
    \item Resource management, network planning, and power control.
    
    \item Received signal strength prediction, interference management, and transmission parameter tuning.
    
    \end{itemize}

    \item \textit{Number of updates}: The training outcome for each round must be transmitted. As a result, much research focuses on lowering the number of update rounds and then the update message itself, i.e., the training derivatives.
        
    \item \textit{Data}: A \acp{tntn} system regularly generates data that is statistically unique from each other due to varied operation, surroundings, or setup devices, i.e. in a non-i.i.d. manner. Because both \ac{cml} and \ac{dml} relies on the i.i.d. assumption, unique strategies to handle statistical heterogeneity must be created.
    
    \item \textit{Simulation models}: Both \ac{cml} and \ac{dml} are data hungry. Since the applications of integrated \acp{tntn} are limited now, it is difficult to get a real dataset. Therefore, simulation models for integrated \acp{tntn} should be defined.
    
    \item \textit{Privacy}: Although \acp{tntn} are designed to protect privacy, it is still necessary to take precautions to ensure that sensitive data is not relegated to specific individuals or devices. Deviating units are the most likely to be harmed since their usage patterns stand out and may influence the model in a unique way.
    
    \item \textit{Dynamism}: The storage, computing, and communication capabilities of a \ac{tntn} system are heterogeneous. As a result, a \ac{tntn} training system must be dynamic or adapt to the device's lowest denominator.

    \item \textit{Theoretical analysis}: The analysis of data driven versus model based methods for integrated \acp{tntn} should be investigated.
    
    \item \textit{Complexity}: Despite the advantages of \ac{cml} and \ac{dml} approaches, most \ac{ml} approaches are computationally heavy. Therefore, these approaches should be investigated in terms of computational complexity, latency, and delay.
    
    \item \textit{Selection of \ac{cml} device}: This critical selection can be based on the device's processing capability, location, scheduling, and memory. Still, research should be done on this problem.

\end{itemize}

\section{Conclusions and Recommendations}
\label{Section5}

\par The \ac{3gpp} has completed \ac{ntn} related studies for \ac{5g} standards and is now working on the \ac{5g}-Advanced standard, which is expected to have fully integrated \acp{tntn}. However, integrating \ac{ntn} devices and networks with the current \ac{tn} technology brings about significant challenges. Much of these challenges are not present in \acp{tn}, and therefore require in-depth studies and novel solutions. Therefore, making \ac{5g} and beyond from space a reality also necessitates initiatives that go beyond standardization. This paper proposes \Ac{ml} approaches for several challenges of the integration of \ac{tntn} and highlights the importance of choosing the appropriate approach, \ac{cml} or \ac{dml}, for various scenarios. Additionally, the feasibility of using these \ac{ml} approaches for each scenario is debated. 

\par Because studies of the application of \ac{cml} and \ac{dml} for \ac{tntn} integration is still recent, this paper also highlights future research areas. Namely, analyses should be made for different use-cases and scenarios of integrated \ac{tntn} for both \ac{cml} and \ac{dml} approaches. Also, algorithms and protocols should be developed to use \ac{ntn} efficiently in real-world applications. Besides that, simulation models, privacy issues, dynamism and so on should be studied in \ac{cml} and \ac{dml} perspectives.

\acrodef{tntn}[TNTN]{terrestrial and non-terrestrial networks}
\acrodef{ntn}[NTN]{non-terrestrial network}
\acrodef{5g}[5G]{fifth-generation}
\acrodef{6g}[6G]{sixth-generation}
\acrodef{ml}[ML]{machine learning}
\acrodef{cml}[CML]{centralized machine learning}
\acrodef{dml}[DML]{decentralized machine learning}
\acrodef{ai}[AI]{artificial intelligence}
\acrodef{kpi}[KPI]{key performance indicator}
\acrodef{haps}[HAPS]{high altitude platform station}
\acrodef{3gpp}[3GPP]{3rd generation partnership project}
\acrodef{fdd}[FDD]{frequency division duplex}
\acrodef{vsat}[VSAT]{very small aperture terminal}
\acrodef{gnss}[GNSS]{global navigation satellite system}
\acrodef{nr}[NR]{new radio}
\acrodef{iot}[IoT]{Internet of things}
\acrodef{geo}[GEO]{geostationary earth orbits}
\acrodef{gso}[GSO]{geosynchronous orbits}
\acrodef{leo}[LEO]{low earth orbit}
\acrodef{tn}[TN]{terrestrial network}
\acrodef{ue}[UE]{user equipment}
\acrodef{rel}[Rel]{Release}
\acrodef{mec}[MEC]{mobile edge computing}
\acrodef{fl}[FL]{federated learning}
\acrodef{bs}[BS]{base station}
\acrodef{meo}[MEO]{medium earth orbit}
\acrodef{hstv}[HSTV]{high speed terrestrial vehicle}
\acrodef{los}[LoS]{line of sight}
\acrodef{nlos}[nLoS]{non-line of sight}
\acrodef{uav}[UAV]{unmanned aerial vehicle}
\acrodef{mac}[MAC]{medium access control}
\acrodef{nb}[NB]{narrowband}

\section*{Acknowledgement}
\label{Section6}
\par This work was supported in part by the Scientific and Technological Research Council of Turkey (TÜBİTAK) under Grant No. 5200030 with the cooperation of Vestel and Istanbul Medipol University.


\end{document}